\documentstyle[12pt]{article}
\def\be{\begin{equation}}
\def\ee{\end{equation}}
\def\l{\label}
\def\ba{\begin{array}}
\def\ea{\end{array}}
\def\ss{\scriptscriptstyle}
\def\Rp{{R\!\!\!\!\!\:/}}

\def\L{{\rm \scriptscriptstyle L \!\!\!\!\!\;/}}

\def\GU{{\rm \scriptscriptstyle GU}}

\def\refe#1{(\ref{#1})}

\def\ltap{\ \raisebox{-.4ex}{\rlap{$\sim$}} \raisebox{.4ex}{$<$}\ }

\def\etal{{\it et al.}}

\def\T{{\it \scriptscriptstyle tripl}}
\def\D{{\it \scriptscriptstyle doubl}}
\def\leff{\lambda_{333}'{}^{\!\!\!\!\!\! \rm \ss eff}}
\def\nuvev{\langle \tilde \nu_3 \rangle}

\def\npb#1#2#3{    {\it Nucl. Phys. }{\bf B #1} (19#2) #3}

\def\plb#1#2#3{    {\it Phys. Lett. }{\bf B #1} (19#2) #3}
\def\prd#1#2#3{    {\it Phys. Rev. }{\bf D #1} (19#2) #3}
\def\prep#1#2#3{   {\it Phys. Rep. }{\bf #1} (19#2) #3}
\def\prl#1#2#3{    {\it Phys. Rev. Lett. }{\bf #1} (19#2) #3}

\def\mpla#1#2#3{   {\it Mod. Phys. Lett. }{\bf A #1} (19#2) #3}

\def\nc#1#2#3{     {\it Nuovo Cim. }{\bf #1} (19#2) #3}

\begin{document}
\begin{titlepage}
\vspace*{-1.5cm}
\begin{center}

\hfill IC/95/122
\\[1ex]  \hfill June, 1995
\\[1ex]  \hfill revised November, 1995

\vspace{5ex}
{\Large \bf Large $R$-parity Violating Couplings
\\[1ex] and Grand Unification}

\vspace{3ex}
{\bf
Alexei Yu. Smirnov$^{a,b}$ and Francesco Vissani$^{a,c}$
}

{\it
\vspace{1ex}  ${}^a$ International Centre for Theoretical Physics, ICTP
\\[-1ex] Via Costiera 11, I-34013 Trieste, Italy
}

{\it
\vspace{1ex}  ${}^b$ Institute for Nuclear Research,
\\[-1ex] Russian Academy of Sciences,
\\[-1ex] 117312 Moscow,  Russia
}

{\it
\vspace{1ex}  ${}^c$ Istituto Nazionale di Fisica Nucleare, INFN
\\[-1ex] Sezione di Trieste, c/o SISSA
\\[-1ex] Via Beirut 2-4, I-34013 Trieste, Italy
}

\vspace{6ex}
{ABSTRACT}
\end{center}

\begin{quotation}
We consider a possibility that
$R$-parity violating interactions
of particles 
which do not involve the first generation
have large (up to 1) coupling constants, $\Lambda$. Such couplings, 
if exist, could have a number of phenomenological consequences:
renormalization of $b-\tau$ mass ratio,
generation of $\nu_\tau$ mass in MeV region, {\em etc.}.
In Grand Unified models, where $B$- and $L$-violating
couplings appear simultaneously, the proton decay can be forbidden
in virtue of hierarchical flavor structure of $\Lambda.$
However, due to Cabibbo-Kobayashi-Maskawa mixing this  decay
is  induced already in one-loop.
Present experimental data give the upper
bound $\Lambda \ltap 10^{-8}$
(or $|\lambda' \lambda''| \ltap 7\cdot 10^{-16},$ on  products
of certain $L$- and $B$-violating coupling constants, in more
general context). The bound can be avoided,
if there is an asymmetry between the $L$- and
$B$-violating couplings of usual matter fields. In the $SU(5)$
model the asymmetry can be related to the
doublet-triplet splitting.

\end{quotation}
\end{titlepage}
\vfill\eject

\section{Introduction}

The gauge invariance of the Standard Model
and Supersymmetry \cite{reviews}
permit, besides usual Yukawa interactions,
\be
\begin{array}{ccl}
 W&=& m_{E,i}/v_1\
    E^c_i (H_1^0 E_i - H^-_1 \nu_i)\\
  &+& m_{D,i}/v_1\
    D^{c\alpha}_i (H_1^0 D_i^\alpha - H^-_1 U_j^\alpha V_{ji}^*)\\
  &+& m_{U,i}/v_2\
    U^{c\alpha}_i (H_2^0 U_i^\alpha - H^+_2 V_{ij} D_j^\alpha)\\
  &+& \mu           (H_1^0 H_2^0 - H^-_1 H^+_2),
\end{array}
\l{R-parity-conserving}
\ee
also the couplings which violate either lepton or
baryon number conservation \cite{WSH}:
\be
\begin{array}{ccl}
 W_\Rp&=& \lambda_{ijk} (  E_i \nu_j - \nu_i E_j ) E_k^c\\
  &+& \lambda_{ijk}' D_i^{c\alpha}
       (\nu_j V_{kl} D_l^\alpha-E_j U_k^\alpha)\\
  &+&  \lambda_{ijk}'' \epsilon_{\alpha\beta\gamma}
                       D_i^{c\alpha}  D_j^{c\beta} U_k^{c\gamma}.
\end{array}
\l{R-parity-violating}
\ee
Here, $E^c_i,E_i,\nu_i,D^c_i,D_i,U^c_i,U_i$ are the superfields
with charged leptons, neutrinos, down- and up-type-quarks;
$i,j,k,l=1,2,3$ are generation indices; $H^0_{1,2},$
$H^-_{1},$ $H^+_{2}$
are the Higgs supermultiplets, and $v_{1,2}$ are the vacuum
expectation values of the scalar components of $H^0_{1,2}.$
The superpotential $W+W_\Rp$
is written in terms of superfields with
fermion mass eigenstates, so that the Cabibbo-Kobayashi-Maskawa
matrix $V_{ij}$ appears in \refe{R-parity-conserving} and
\refe{R-parity-violating} explicitly;
$m_{E,i},$ $m_{D,i},$ $m_{U,i}$
are the fermion masses.
Another possible term in $W_\Rp$,
 $\mu_i (\nu_i H_2^0 - E_i H^+_2)$,  can be
rotated away from the superpotential, by  redefinition
of the  couplings in $W$ and $W_\Rp$ .

A rich phenomenology can be related to the
interactions \refe{R-parity-violating}.
They result in  $B$- or/and $L$-violating phenomena like
$n-\bar n$ oscillations
\cite{Zwirner,Barbieri-Masiero,Goity-Sher},
proton decay \cite{HN,pd},
generation of Majorana neutrino masses
\cite{neutrino-masses1,neutrino-masses2},
neutrinoless double beta decays
\cite{0nu-bb1,0nu-bb2};
they  modify usual processes
like $\mu$-, $\beta$-decay \cite{Barger-Giudice-Han},
and lead to the decay of the
lightest supersymmetric particle \cite{lsp-decay}.
However,
up to now no effects of
\refe{R-parity-violating}
have been found  which implies strong restrictions on
the constants $\lambda.$
In particular the proton decay searches allow to put
the bound on certain couplings of
lowest generations:
\be
|\lambda'\lambda''| \ltap 10^{-24}
\l{proton-decay-bound-on-lambda}
\ee
for squark masses around 1 TeV.

The smallness of at least some couplings  \refe{R-parity-violating}
indicates that probably all
the interactions \refe{R-parity-violating} are absent in virtue of
certain symmetry.
Moreover, the absence of the terms \refe{R-parity-violating}
ensures stability of the lightest supersymmetric particle
which is considered as a favorite  candidate for the cold dark matter.
$W_\Rp$ can be suppressed by $R$-parity
or matter parity conservation. The
corresponding  symmetries may  naturally follow
from a class of Grand Unified symmetries like $SO(10)$
in models with minimal particle content.
Alternatively $B$- or $L$-violating
terms can be suppressed by symmetries which distinguish
quarks and leptons.

In this paper we assume that $R$-parity (or some other
symmetry which suppresses $W_\Rp$)
{\em is not exact} and the terms \refe{R-parity-violating}
are
generated with sufficiently small coupling constants.
In fact, the existing data strongly restrict the
couplings of light generations,
whereas the bounds on couplings of second and
third generations are weak or absent
(for latest discussion see \cite{CRS}).
In the same time it is natural to assume the
hierarchy of constants $\lambda$ \cite{hierarchy}.
Moreover, as the consequence of a horizontal symmetry,
this hierarchy
can be much stronger than that of the usual Yukawa couplings.
Strong hierarchy of $\lambda$ can be partially related to the fact that
couplings in
\refe{R-parity-violating} involve three generation dependent fields,
whereas Yukawa couplings
contain  only two such fields
(see for latest discussion \cite{CRS}).
Thus the following pattern is possible:
the constants $\lambda$ for the first and second
generations are very small and satisfy
the existing bounds, while the couplings
involving third generation particles are large and could be
of the order 1.

Large $R$-parity violating couplings  of third generation
can manifest themselves in many ways.\\
At one-loop they
induce the Majorana neutrino masses
\cite{neutrino-masses1,neutrino-masses2}.
They contribute to $K^0-\bar K^0$
mixing, to the electric dipole of the neutron
\cite{Barbieri-Masiero},
to $Z\to b\bar b$ decay width \cite{Bh},
the decay of $B$ meson $B^-\to K^0 K^-$ \cite{CRS} {\em etc.}.\\
Large $\lambda '$s influence the running of usual
Yukawa couplings. In particular, they
modify the infrared fixed point of the top quark
Yukawa coupling \cite{Biswajoy}. The restriction
$\lambda''_{i33}<0.4-0.5$
has been obtained from the condition
that the top coupling does not blow up
before the Grand Unification  scale $M_\GU$.
Large $B$- or $L$-violating couplings
of the heaviest generations
can appreciably renormalize the $b-\tau$ mass ratio.
It is shown \cite{prep}
that for values $\lambda''_{233} = 0.15 - 0.30$ the
$(b-\tau)$-mass unification at GU scale can be achieved for
any value of $\tan\beta$ in the interval $2-50$.

The studies of the $R$-parity violation effects were
performed mainly in the context of Minimal Supersymmetric
Standard Model.
However remarkable convergency of
the gauge couplings at the scale around $3 \cdot 10^{16}$ GeV
\cite{GCC1,GCC2,GCC3}
can be considered as  strong indication of the
supersymmetric unification of the strong and the electroweak interactions.
Supersymmetry offers an elegant way to stabilize the gauge
hierarchy, thus ensuring consistency of the picture.
Moreover, the $b-\tau$ unification \cite{btau}
can be achieved
in the supersymmetric GU model only \cite{btau-in-susy}.
Note that $\lambda$-couplings, like the usual Yukawa couplings,
will affect only weakly (at the two-loop level)
the evolution of the gauge coupling constants.
In this connection it is important to consider
the properties and consequences of the interactions
\refe{R-parity-violating}
in the GU theories.
The first studies of $R$-parity violation in the context of
Grand Unification have been performed in
\cite{Ramond,Hall-Suzuki,Brahm-Hall}.

In this paper we consider the proton decay induced by $R$-parity
violating couplings of heaviest (second and third) matter generations.
We find new very strong bounds on $\lambda$ in the $SU(5)$
with standard matter field content.
The modifications of the model are discussed which allow us
to get the asymmetry of $B$- and $L$-violating couplings
and thus to avoid the bounds.

The paper is organized as follows. Properties of $R$-parity violating
couplings in SUSY $SU(5)$ are discussed in sect.\ 2.
We consider the proton
decay induced by these couplings
in sect.\ 3. The conditions are found at which the decay is forbidden
in the lowest  order of perturbation theory. However,
being suppressed in lowest order, proton decay  is inevitably generated
by one-loop diagrams (sect.\ 4). The amplitudes of leading  one-loop
diagrams are estimated and the upper bounds on $R$-parity
violating coupling constants are found. In sect.\ 5 we consider the
generality of the bounds and the way to avoid them. Then (sect.\ 6)
we discuss possible relations between asymmetry of the
$B$- and $L$-violating interactions which allows
one to avoid the bounds and the doublet-triplet splitting. Sect.\ 7
summarizes the results.

\section{$R$-parity violating interactions in the
$SU(5)$-su\-per\-sym\-me\-tric model.}

In the $SU(5)$ model one can introduce the following
$R$-parity violating interactions \cite{Ramond}
\be
\Lambda_{ijk} \bar 5_i \bar 5_j 10_{k} +
\bar 5_i(M_i + h_i \Phi) H,
\l{lambda-su5}
\l{R-viol-su5}
\ee
where $i,j,k=1,2,3$ are generation indices,
$\Lambda_{ijk}$ are the coupling constants and
$\bar 5_i,$ $10_i$ are the matter superfields which
can be written in terms of the standard model
supermultiplets as:
\be
\bar 5 = \left(
               \begin{array}{c}
                   D^{c}\\
                   i\sigma_2 L
               \end{array}
           \right)
\ \ \ \ \ \ \ \ \ \
10 = \left(
               \begin{array}{cc}
                   U^{c} & -Q\\
                   Q  & -E^c i\sigma_2
               \end{array}
      \right).
\l{schematic-notation}
\ee
Here $\sigma_2$ is the Pauli matrix,
$L=(\nu,E)$ and $Q=(U,D)$ are $SU(2)_L$ doublets.
$M_i$ are the mass parameters, $h_i$ are  couplings,
$\Phi$ and $H$ are the 5-plet and  24-plet
of Higgs fields.

Let us consider first the effects of
$\Lambda$ couplings, suggesting that
the matter-Higgs mixing (second term in \refe{R-viol-su5}) is
negligibly small.
The $\Lambda_{ijk}$-couplings
\refe{R-viol-su5} generate all the $R$-parity violating interactions (2).
It is convenient to define
$\Lambda_{ijk}$ in the basis,
where $SU(2)_L$-singlets $u^c$ and $d^c$
(fermionic components of $U^c$ and $D^c$)
coincide with mass eigenstates. This always can be done since
$u^c$ and $d^c$ enter different $SU(5)$-multiplets.
Note that due to the antisymmetry of
10-plets the interactions \refe{R-viol-su5} are antisymmetric
in generation indices: $\Lambda_{ijk}=-\Lambda_{jik}.$

Substituting multiplets \refe{schematic-notation} in
\refe{R-viol-su5}
and comparing the resulting interactions
with those in
\refe{R-parity-violating} we find the relations between
original $\lambda_{ijk}$ and $\Lambda_{ijk}$ couplings
at the GU scale:
\be
\begin{array}{l}
\lambda_{ijk}=\Lambda_{ijl} V_{lk}\\
\lambda'_{ijk}=2\Lambda_{ijk}\\
\lambda''_{ijk}=\Lambda_{ijk}.
\l{lambda-unified1}
\end{array}
\ee
As a consequence of
quark and lepton unification in $SU(5),$ all types of $R$-parity
violating couplings
appear simultaneously. Moreover, different couplings
$\lambda,\lambda'$ and $\lambda''$ are determined by unique
GU coupling $\Lambda.$ As follows from
\refe{lambda-unified1}, up to CKM matrix and
factor 2 in $\lambda'$ these couplings coincide at GU scale:
\begin{equation}
\lambda_{ijl} V^{-1}_{lk}=\frac{1}{2} \lambda'_{ijk}=\lambda''_{ijk}.
\l{lambda-unified2}
\end{equation}
Evidently, there is no relative suppression of $B$- and
$L$-violating couplings.
Another feature of the  Grand Unification is that $L$-violating
couplings, $\lambda'_{ijk},$
should be antisymmetric in first two indices:
$\lambda'_{ijk}=-\lambda'_{jik},$ similarly
to other couplings. In the non-unified version
\refe{R-parity-violating} these couplings can have
also a symmetric part.

The gauge coupling renormalization effects
lead to modification of GU relations \refe{lambda-unified1}
at the electroweak scale:
\be
\begin{array}{l}
\lambda_{ijk}=1.5\ \Lambda_{ijl} V_{lk}\\
\lambda'_{ijk}=2\ (3.4 \pm 0.3)\ \Lambda_{ijk}\\
\lambda''_{ijk}=(4.4 \pm 0.4)\ \Lambda_{ijk},
\l{lambda-running}
\end{array}
\ee
where the errors correspond to the uncertainty in strong coupling
constant:
$\alpha_s(M_Z)=0.12\pm 0.01.$
Inclusion of other
uncertainties related {\it e.g.}\  to threshold
SUSY and GU corrections
may require the doubling of the errors quoted.
The renormalization effects due to third family
Yukawa couplings \cite{prep}
do not drastically change the relations
\refe{lambda-running}.
Let us define the renormalization factor $\eta/2,$
relevant for proton decay as:
\begin{equation}
\lambda'(M_Z)\lambda''(M_Z)=\eta\cdot \Lambda^2.
\l{eta}
\end{equation}
{}From equation \refe{lambda-running} we find: $\eta=30\pm 5.$

\section{Proton decay due to
$R$-parity violating couplings in the lowest order.}

Simultaneous presence of both $B$- and $L$-violating
couplings in GU models leads to proton decay.
Let us consider the proton decay
taking into account GU relations between couplings
\refe{lambda-unified1}.
There are two types of decay modes:

(1) $(B-L)$-conserving decays. The exchange of $\tilde d^c_i$
squarks between $B$-violating and $L$-violating vertices
induces the 4-fermion operators:
\be
2\eta \frac{\Lambda_{ijk}^* \Lambda_{ilm}}{m^2_{\tilde d_i}}\,\,\,\,
\overline{d_j^c u_k^c}\ (V_{mn} d_n \nu_l  - u_m e_l ).
 \l{induced-b-l}
\ee
The kinematics selects the following 4-fermions operators
in \refe{induced-b-l}
\begin{equation}
\begin{array}{l}
\overline{d^c u^c}\ \nu d, \,\,\,\
\overline{d^c u^c}\ \nu s, \,\,\,\
\overline{d^c u^c}\  e u , \,\,\,\
\overline{d^c u^c}\ \mu u,\\
\overline{s^c u^c}\ \nu d, \,\,\,\
\overline{s^c u^c}\ e u, \,\,\,\
\overline{s^c u^c}\ \mu u
\end{array}
\end{equation}
which lead to the proton decay.
All these operators contain the $u^c$ quarks, and therefore
can be forbidden at tree level,
if in the  basis where $u^c_{i}$ are the mass eigenstates
we put
\be
\Lambda_{ij1}=0.
\l{b-l-forbidden}
\ee

(2) $(B+L)$-conserving decays. The mixing of squarks:
$\tilde b^c,\ \tilde b $ and
$\tilde t^c,\ \tilde t$ leads to the
operators:
\be
\begin{array}{l}
\displaystyle
2\eta \frac{(\Lambda_{3jk} \Lambda_{lmn} V_{n3})^*}
{{\cal M}^2_{\tilde b}}\
\overline{d_j^c u_k^c}\ \overline{d_l^c \nu_m}\ , \\
\displaystyle
-\eta \frac{(\Lambda_{ij3} \Lambda_{lm3} )^*}{{\cal M}^2_{\tilde t}}\
\overline{d_i^c d_j^c}\ \overline{d_l^c e_m}\ ,
\end{array}
\l{induced-b+l}
\ee
where ${{\cal M}^2_{\tilde b}}$ and
${{\cal M}^2_{\tilde t}}$ parametrize the
propagators $\tilde b^c - \tilde b$
and $\tilde t^c -  \tilde t$
for low momenta. In particular,
\be
\frac{1}
{{\cal M}^2_{\tilde b}} =
\frac{m^2_{\tilde{b} LR}}
{m^2_{\tilde b LL} m^2_{\tilde b RR} - m^4_{\tilde b LR}},
\ee
where the mixing parameter ${m^2_{\tilde b LR}}$
is induced both via the $\mu$-term
at SUSY conserved level
and via the soft breaking terms:
\be
m^2_{\tilde b LR} = m_b (A_b + \mu \tan \beta).
\l{LR-mixing}
\ee
Here $A_b = O(m_{3/2})$ is soft breaking parameter.
For $\tan \beta \sim 20 - 50$ the mixing mass
may not be suppressed with respect to
the diagonal masses
$m^2_{\tilde b LL}$ and $m^2_{\tilde b RR}.$
Consequently, the propagator factor
${1}/{{\cal M}^2_{\tilde b}}$
as well as ${1}/{{\cal M}^2_{\tilde t}}$
can be of the  order of the factor $1/m^2_{\tilde d_i}$
from Eq.\ \refe{induced-b-l}.
We neglect the mixing of squarks from the lightest generations
which are proportional to light quark masses.
Mixing between squarks of different generation
should be negligibly small
to avoid the constraints from non-observation of
flavor changing neutral-currents.
Taking into account the kinematics we find from
\refe{induced-b+l} the operators
leading to proton decay:
\be
\overline{ d^c u^c}\ \overline{ d^c \nu},\,\,\,\
\overline{ s^c u^c}\ \overline{ d^c \nu},\,\,\,\
\overline{ d^c u^c}\ \overline{ s^c \nu},\,\,\,\
\overline{ d^c s^c}\ \overline{ d^c \mu}.
\ee
The first three operators (with $u^c$)
disappear if the conditions  \refe{b-l-forbidden} are fulfilled;
the last one can  be removed by the equality
\be
\Lambda_{123}=0.
\l{b+l-forbidden}
\ee

In fact, $\Lambda_{ij1}$ and $\Lambda_{123}$ may not
be precisely zero;
using relations \refe{lambda-unified2}
and renormalization effect \refe{eta}
we get from
\refe{proton-decay-bound-on-lambda}
the bound on the GU scale couplings:
\begin{equation}
\Lambda_{ij1},\ \ \Lambda_{123} \ltap 2\cdot 10^{-13}.
\l{tree-level-bound}
\end{equation}
In both conditions \refe{b-l-forbidden} and \refe{b+l-forbidden}
the coupling constants with first family index are involved.
Therefore we can assume the family hierarchy, according
to which the couplings with low indices ({\it i.e.}\ 1 and 2)
are small, and maximal couplings are those with maximal
number of family indices 3,
first of all $\Lambda_{233},$ and then, probably, $\Lambda_{133}.$
The question is: How large can be $\Lambda_{233}$?

\section{Proton decay induced by $\Lambda_{233}$ at one-loop.}

Let us consider the configuration being the most protected from the
proton decay, when there is only one term,
$\Lambda_{233}\bar 5_2 \bar 5_3 10_3,$
in \refe{R-viol-su5},
with the following fermionic content of the supermultiplets:
$10_3$, includes  $t^c$,
$q_3 = (t, b'),$ where $b'\equiv V_{3i} d_{i},$
and $\tau^c;$
$\bar 5_3$ contains $b^c$ and  $l_{3}=(\nu_\tau,\tau),$
$\bar 5_2$ contains
$s^c$ and  $l_{2}=(\nu_\mu, \mu).$
All other terms in \refe{R-viol-su5} have  zero or
negligibly small couplings.
We will show that even in this case the proton decay appears
as one-loop effect, thus leading to still very strong bounds
on $\Lambda_{233}$.

Note that there is only one $B$-violating vertex:
$\Lambda_{233}D^c_2 D^c_3 U^c_3.$
It can be connected to $L$-violating
vertex (needed to proton decay) by exchange of $b^c$, $s^c$
or $t^c$ and  corresponding squarks.
This allows to systematically find all relevant
diagrams for proton decay.
In accordance with
\refe{induced-b-l} and \refe{induced-b+l} we get
at the tree level
the following $(B-L)$-conserving
\be
\displaystyle
2\eta{|\Lambda_{233}|^2}
\left[\frac{1}{m^2_{\tilde s}}
\overline{b^c t^c}\  (b'\nu_\tau-t \tau)
+
\frac{1}{m^2_{\tilde b}}
\overline{s^c t^c}\  (b' \nu_\tau-t \mu)
\right]
\l{lambda233-b-l}
\ee
and the $(B+L)$-conserving:
\be
\displaystyle
2\eta(\Lambda_{233}^*)^2
\left[
\frac{V_{tb}^*}{{\cal M}^2_{\tilde b}}
\overline{s^c t^c}\ (\overline{b^c \nu_\mu}
-\overline{s^c \nu_\tau})
+
\frac{1}{{\cal M}^2_{\tilde t}}
\overline{s^c b^c}\ (\overline{b^c \mu}
-\overline{s^c \tau})
\right]
\l{lambda233-b+l}
\ee
operators.
Also the operators are generated which
can be obtained from \refe{lambda233-b-l} and
\refe{lambda233-b+l} by replacement of two
ordinary particles by their superpartners.
The terms with
$\tilde s^c - \tilde s$
exchange are omitted
in \refe{lambda233-b+l}, since
they are proportional to small factor
$V_{ts}/
{{\cal M}^2_{\tilde s}}
\approx 10^{-3} /
{{\cal M}^2_{\tilde b}}
$
(in this equality we took into account that
$\tilde s^c - \tilde s$ mixing is suppressed with respect to
$\tilde b^c - \tilde b$ mixing by  $m_s/m_b,$
see \refe{LR-mixing}, and
therefore
${{\cal M}^2_{\tilde b}}/
{{\cal M}^2_{\tilde s}} \sim m_s/m_b$).\vskip0.5truecm

As we discussed before
for kinematical
reasons the operators \refe{lambda233-b-l} and \refe{lambda233-b+l}
do not lead to proton decay.
However, an additional exchange by the $W$-boson (or wino)
as well as by  charged Higgs (or Higgsino) converts
the operators  \refe{lambda233-b-l}\ and \refe{lambda233-b+l}
(or the operators with superpartners)
to the operators with light fermions which give
proton decay already at one-loop level.
Indeed, due to the presence of the CKM mixing
the $W$- (wino), charged Higgses (Higgsino) have family off-diagonal
couplings
(see Eq.\ \refe{R-parity-conserving}).
The emission or absorption
of these particles can reduce the generation index.

Let us
find, using the
operators \refe{lambda233-b-l} and \refe{lambda233-b+l},
the crucial factors which appear in such a
generation reduction:\\
(i). Evidently, the second term of
\refe{lambda233-b-l} with  four heavy fermions, and
the fourth term with  two $t$ quarks, can not be
transformed at one-loop into the  operators with light particles
only.
Similarly, the third and the fourth  terms in \refe{lambda233-b+l}
stipulated by
$\tilde t^c - \tilde t$ exchange do not give $p$-decay at one-loop.
The third
term contains two $b^c$ quarks, the fourth one has three heavy
fermions ($m > m_p$).\\
(ii). All the rest operators include $t^c$ ($\tilde{t}^c$).
The $t^c \rightarrow d$ conversion
due to emission of charged Higgs boson
or $W$-boson gives the factor
$V_{td}\ m_t$
(in the case of the Higgs this factor follows from the Yukawa coupling
\refe{R-parity-conserving};
in the case of the $W$-exchange it comes from the chirality flip:
$t^c \rightarrow t \rightarrow d\ W$). The same factor appears
for $\tilde {t}^c \rightarrow d$ transition. Similarly, the conversion
${t}^c \rightarrow s$ ($\tilde{t}^c \rightarrow s$) implies the factor
$V_{ts} m_t$.\\
(iii). The amplitudes of transitions  of down quarks (squarks)
$b^c \rightarrow u$
($\tilde{b}^c \rightarrow u$), and
$s^c \rightarrow u$
($\tilde{s}^c \rightarrow u$) are proportional, respectively,
to $V_{ub}^*\ m_b$ and  $V_{us}^*\ m_s$.
These factors are of the same order of magnitude.\\
(iv). $L$-violating part of $(B-L)$-operators
\refe{lambda233-b-l} contains
small CKM-elements $V_{ts}$ or $V_{td}$, whereas
$(B+L)$ operators \refe{lambda233-b+l}
are proportional to $V_{tb} \approx 1$.

Combining the factors discussed in (ii)-(iv)
we find  that the largest one-loop amplitudes of $(B-L)$- and
$(B+L)$-conserving  modes contain the additional loop factors
\begin{equation}
\begin{array}{l}
\displaystyle
\xi_{B-L} = \frac{m_b m_t}
{16 \pi^2 v^2}
V_{ub}^* V_{td} V_{ts} \nonumber\\[1ex]
\displaystyle
\xi_{B+L} = \frac{m_b m_t}
{16 \pi^2 v^2}
V_{ub}^* V_{td} V_{tb}^*,
\l{xi}
\end{array}
\l{xi-factors}
\end{equation}
where
$v\equiv \sqrt{v_1^2+v_2^2},$ and $1/16 \pi^2$
comes from loop integration.
There are also transitions for which the loop factors
can be obtained from
\refe{xi} by  substitution
$V_{ub}\ m_b \rightarrow V_{us}\ m_s$.

The second and the first terms in \refe{lambda233-b-l} as well as the third 
and the fourth terms in \refe{lambda233-b-l} can be transformed 
to the operators with 
light particles which induce the proton 
decay by additional exchange 
of $W$ ($\tilde{W}$) or $H$ ($\tilde{H}$). 
Thus the proton decay will 
appear at two-loop level. 
An additional (to \refe{xi-factors}) suppression factor 
for two-loop diagrams is $\sim (g^2/16 \pi^2) \sim 10^{-3}$. 
\vskip0.5truecm

Let us estimate the contributions from the leading diagrams.

1. The propagation of squark $\tilde b^c$
between $B$- and $L$-violating vertices
(first term in \refe{lambda233-b-l})
``dressed" by charged Higgs (Higgsino) interaction leads to the diagrams
shown in Fig. 1a,b. The mixing of charged Higgses $H^-_1$, $H^+_2$ in
diagram of Fig. 1a and the coupling of three squarks in Fig.\ 1b
are induced by soft
SUSY breaking terms $\mu B H_1 H_2$ and $A \Lambda_{233}
\tilde s^c \tilde b^c  \tilde t^c  $, correspondingly, where
$A, B = O(m_{3/2})$. The estimation of diagrams gives
\begin{equation}
\displaystyle
2\eta \frac{|\Lambda_{233}|^2}{m^2_{\tilde{b}}}
\xi_{B-L}
\left[
d u\  s \nu_{\mu}+s u\ d \nu_{\mu}
\right].
\l{loop-b-l1}
\end{equation}
In \refe{loop-b-l1}
we have taken into account the relation
\begin{equation}
\frac{\mu A}{v_1 v_2 m_{H^+}^2} \sim
\frac{\mu B}{v_1 v_2 m_{H^+}^2} =
\frac{1}{v^2},
\l{relation}
\end{equation}
that connects
the mass of the physical charged Higgs,  $m_{H^+}^2,$
and
the parameter
of the mixing of scalar doublets $H_1$ and $H_2,$
$\mu B.$
The diagrams with dressing by $W$ and wino (Fig.\ 1c,d) give  similar
result.

2. There are also the box diagrams with $\tilde{b}^c$ exchange, when
$H^+$ emitted by $t^c$ is absorbed by quark $b'$ from $L$-violating
vertex. Since $b' \rightarrow u^c$
transition is forbidden ($b'$ couples to $t^c,$ or $t$) the diagram
gets the GIM suppression factor
$V_{ub} m_b^2/\tilde{m}^2$, where $\tilde{m}^2$
is typical mass of squark. As the result one gets
$$
\frac{A_{box}}{A_{vertex}} \propto \frac{m_u}{V_{ts} m_b}
\frac{m_b^2}{\tilde{m}^2} < 10^{-4}.
$$
Box diagrams lead to  $(V-A)$-Lorentz structure of the
effective operators.

3. The exchange of $\tilde{s}^c$ squark gives the diagram similar to those
in Fig.\ 1a-d with emission of
$\bar{\nu}_{\tau}$ instead of $\bar{\nu}_{\mu}$.
The amplitude can be obtained from \refe{loop-b-l1} by substitution
\begin{equation}
\frac{{m_b} V_{ub}^*}{{m_{\tilde{s}}^2}}
\rightarrow
\frac{{m_s} V_{us}^*}{m_{\tilde{b}}^2} .
\l{substitution}
\end{equation}

4. ``Dressing" the $(B+L)$ diagram with
$\tilde{b}^c - \tilde{b}$ exchange
(first and second terms in \refe{lambda233-b+l})
by $H^{\pm}$ and $\tilde H^{\pm}$ one gets
the diagrams shown in Fig.\ 2a-d. Similar diagrams exist with dressing by
$W$-boson and wino. The amplitudes
corresponding to Fig. 2a,b
can be estimated as:
\begin{equation}
2\eta
\frac{(\Lambda_{233}^*)^2}
{{\cal M}^2_{\tilde{b}}}
\xi_{B+L}\ d u\ \overline{s^c \nu_\tau}.
\l{loop-b+l1}
\end{equation}

5. Box diagrams shown in Fig. 2c,d and similar diagrams with $W$ and
$\tilde {W}$ give the amplitudes comparable with that in
\refe{loop-b+l1}
but having $(V - A)$-structure:
\begin{equation}
2\eta
\frac{(\Lambda_{233}^2)^*}{{\cal M}^2_{\tilde{b}}}
\xi_{B+L}\ 
\bar{s^c}\bar \sigma^{\alpha} d \ \bar{\nu}_{\mu} \bar \sigma_{\alpha} u.
\l{loop-b+l2}
\end{equation}

6. The contributions of diagrams with exchange of
$\tilde{s}^c-\tilde{s}$  (similar
to those in Fig.\ 2a,b) are suppressed
by factor of  $m_s/m_b V_{ts}$, as
we marked before.\vskip0.5truecm

According to previous discussion,
the ratio of
$(B+L)$- and $(B-L)$-amplitudes is
\be
\frac{A_{B+L}}{A_{B-L}} \sim
\frac{1}{V_{ts}}
\frac{m_{\tilde{b} LR}^2}{m_{\tilde{b}}^2} .
\ee
For large tan$\beta$ one has
$m_{\tilde{b} LR}^2 \sim m_{\tilde{b}}^2$ , and therefore $(B+L)$
amplitudes are enhanced by factor $1/V_{ts}$.
Consequently,  in models with $\Lambda_{233}$ being  the main
source of $R$-parity violation
the decay channels $p\to K^+ \nu_\tau$
and $p\to K^+ \nu_\mu$ dominate over
$p\to \pi^+ \overline{\nu_\tau},$
$p\to K^+ \overline{\nu_\tau}$
channels (and similar modes with
$\nu_\mu$). The $(B-L)$ channels may have branching ratios as
small as  $|V_{ts}|^2\sim 10^{-3}.$
In the case of  $\tan \beta \sim 1$ the $(B-L)$- and
$(B+L)$-amplitudes can be of the same order of magnitude.

Thus proton decay forbidden in the lowest order
is generated  due to
CKM-mixing in  one-loop. As follows from \refe{loop-b-l1},\
\refe{loop-b+l1}\ and \refe{loop-b+l2}
an additional suppression factor \refe{xi} appears in one
loop amplitudes in comparison with tree level ones.
Numerically it equals
\be
\xi_{B+L} \equiv \xi = 5\cdot 10^{-9}
\left(\frac{m_b}{4.6\ {\rm GeV}}\right)
\left(\frac{m_t}{176\ {\rm GeV}}\right)
\left(\frac{V_{ub}^*}{3\cdot10^{-3}}\right)
\left(\frac{V_{td}}{ 10^{-2}}\right).
\ee
Consequently,  the bound
on $\Lambda_{233}$ can be relaxed
by factor $\sqrt {\xi} \approx 7 \cdot 10^{-5}$:
\be
\Lambda_{233} \ltap 3 \cdot 10^{-9}
\l{cons-bound}
\ee
(compare with \refe{tree-level-bound}).
Using the amplitude \refe{loop-b+l1}
which can dominate at large $\tan \beta$
we find:
\begin{equation}
\Lambda_{233}^2 \ltap
8\cdot 10^{-18} \left( \frac{10^{-2}}{V_{td}} \right)
\left(
\frac{{\cal M}^2_{\tilde b}}{1{\rm TeV}^2}
\right).
\l{bound}
\end{equation}
This result coincides with \refe{cons-bound}
at $V_{td} \sim 10^{-2}$
and ${{\cal M}^2_{\tilde b}} \sim 1$ TeV.

Thus bounds on the proton lifetime  strongly restrict  even  the
$\Lambda_{233}$ coupling of  highest generations of matter
fields. Large $R$-parity violating coupling constants are not admitted
for any generation.\vskip0.5truecm

The following remarks are in order.\\
1). $V_{td}$ is a common coefficient of all the amplitudes.
For $V_{td}=0$ one might have  the suppression of all the one-loop
contributions. However, the unitarity constraints
of the CKM matrix give for
$V_{td}=(1\pm 0.5)\cdot 10^{-2}$ at 90\% C.L.\\
2). Lorentz structure of the one-loop operators
differs from that of the tree level operators.
In particular, the vertex diagrams result in change of chirality
(from right to left) of quarks from $B$-violating couplings.
In box diagrams there is a change of chirality of one
quark from $B$-violating
and one quark from $L$-violating couplings.
Therefore no cancellation between one-loop and tree level
contributions is expected.\\
3). The explicit computations of the diagrams confirm
the results
\refe{loop-b-l1},\
\refe{loop-b+l1}\ and \refe{loop-b+l2},
up to the factor
\be
\frac{\ln x}{x-1},
\l{loop-function}
\ee
where $x \equiv m_t^2/m_{H^+}^2$ for Higgs dressing,
and  $x \equiv m_t^2/m_{W}^2$ for $W$-dressing.
$W$-contributions have also an additional factor 3.
For $m_{H^+}>250$ GeV the contributions from Higgs dressed
diagrams exceed those from diagrams with $W$.\\
4). Due to the relation \refe{relation}
there is no dependence of amplitudes on
$\mu B,$ $m_{H^+}^2$ or $\tan\beta$.
This result is confirmed by explicit computation of diagrams up to the
above factor \refe{loop-function}.\\
5). Analysis performed in this section for
$\Lambda_{233}$
is  valid for all couplings
$\Lambda_{ijk}$ which do not result in $p$-decay in the
lowest order.
For other couplings the bounds are of the order of 
\refe{cons-bound} 
or 
even stronger. Let us consider  
$\Lambda_{232}$,  
another coupling which does not contain the first generation 
index. Now $c^c$ quark enters  baryon violating coupling 
and in the  amplitudes found above one should substitute 
$m_t V_{td}$ by $m_c V_{cd}$. The latter product 
is about 13 times smaller than the former one. 
Consequently, 
$(B+L)$ amplitudes will be suppressed by additional factor 13 
and therefore the bound on  
$\Lambda_{232}$ 
will be relaxed (if there is no cancellation 
of contributions from different diagrams) 
in comparison with 
\refe{cons-bound}:  $\Lambda_{232} < 10^{-8}$.  
This bound  
can be considered as the conservative
bound on all $R$-parity violating coupling.\\
6). The analysis performed above and the  bounds on $R$-parity
violating constants are valid in more
general context  without Grand Unification.
In (27) $\Lambda_{233}^2$ should be substituted by the product
$|\lambda'_{233} \lambda''_{233}|$.
Taking into account the
renormalization effects we get at the electroweak scale:
\begin{equation}
|\lambda'_{233} \lambda''_{233}| \ltap
5\cdot 10^{-17} \left( \frac{10^{-2}}{V_{td}} \right)
\left(
\frac{{\cal M}^2_{\tilde b}}{1{\rm TeV}^2}
\right).
\end{equation}
Similar or even stronger bounds can be obtained for the 
products of 
$\lambda'$ and $\lambda''$ couplings which can reproduce the 
tree level diagrams of the type shown in Fig.\ 1, 2 (without 
dressing). Namely, the results of this section  
can be immediately applied  to 
$|\lambda'_{iab} \lambda''_{icd}|$, 
$|\lambda'_{iab} \lambda''_{cid}|$ ($(B - L)$ modes) and  
$|\lambda'_{abi} \lambda''_{icd}|$, 
$|\lambda'_{abi} \lambda''_{cid}|$ ($(B+L)$ modes). 
The corresponding diagrams for any values of 
indices ($a, b, c, d = 1, 2, 3$) lead to proton decay either at 
tree level, or after ``dressing" at one or two-loop level. 
The combination 
$|\lambda'_{iab} \lambda''_{cdi}|$ allows one to construct 
the following diagram: the (s)quark $U_i^c$ emitted from 
baryon violating vertex is converted to $\tilde{d}_i$ squark by 
interaction with charged Higgs (Higgsino). 
Then $\tilde{d}_i$ is transformed 
to $\tilde{d}^c_i$ by mixing mass term and the latter 
is absorbed in lepton violating vertex with coupling 
$\lambda'_{iab}$. It can be shown that such 
a type of diagrams can be constructed for 
any combination of   
$\lambda'$ and $\lambda''$ couplings. Being dressed by 
Higgs/$W$-loops they lead to proton decay. 
This means that any such a combination can be restricted 
by proton decay data at some level \cite{prep}.\\ 
7). The presence of matter-Higgs mixing terms \refe{R-viol-su5}
does not change the bounds \refe{cons-bound} unless strong
fine tuning is implied.

\section{Can $R$-parity violating couplings be large?}

In previous section we have considered the model with MSSM 
particle content. It has been shown that 
even very strong hierarchy of $R$-parity 
violating couplings (such that all but 
$\Lambda_{233}$ can be neglected) does not allow to get  
$\Lambda_{233} \sim 1$. In what follows  
we turn down this minimality 
admitting an  existence of additional Higgs 
or/and matter superfields.  Also we will not rely 
of family hierarchy of $R$-parity violating couplings, considering 
the most general case.  
To what extend this allows one to relax the 
bound \refe{cons-bound}?\vskip0.5truecm

We start by possible effects of the extended Higgs sector. 
In the case of  complex Higgs sector
({\it e.g.}\ with additional 45-plets)
there is a possibility to make another arrangement
of particles in the $SU(5)$ multiplets.
(In fact, such a sector allows to reproduce correct
mass ratios $m_e/m_d,$ $m_\mu/m_s$).
In principle, an arbitrary mixing (permutations)  of the
$SU(2) \times U(1)$ blocks from  5-plets
as well as 10-plets of different generations
are admitted. In particular, in
$\bar{5}_3,$ together
with $b^c$-quark it is possible to put some
combination of the leptonic doublets:
\be
L_3 \to (\hat U L)_3 \equiv \hat U_{3i} L_i,
\l{rl}
\ee
and together with $t^c$ in $10_3$ one can put some combination
of quark doublets:
\be
Q_3 \to (\hat W Q)_3 \equiv \hat W_{3j} Q_j,
\l{rq}
\ee
where $\hat U$ and $\hat W$ are arbitrary unitary matrices.
Such a mixing of the
$SU(2) \times U(1)$ blocks changes the structure of $R$-parity
violating couplings, modifying the relation
\refe{lambda-unified2}.
In particular,
for $\lambda'$ and $\lambda''$ we get
\be
\lambda'_{ijk} = 2 \lambda''_{ij'k'} \hat U_{j'j} \hat W_{k'k}
\ee
instead of \refe{lambda-unified2}.

We prove in the following that the freedom given by
the rotations \refe{rl} and \refe{rq} is
not sufficient to avoid proton decay at one-loop level.
Suppose again that only
$\lambda_{233}''$ is non-zero. Dressing of the vertex
$\lambda_{233}'' (\tilde s^c b^c t^c -\tilde b^c s^c t^c)$
by Higgs (Higgsino) gives in one
loop:
\be
\lambda''_{233} \xi \left (\tilde{s}^c -  k \tilde{b}^c \right)
\left(\frac{V_{td}}{V_{ts}} u d + u s \right),
\l{B-vertex}
\ee
where $k$ is a constant of the order 1,
and $\xi$ is the one  loop
suppression factor \refe{xi}.
The coupling (35)  does not depend on $\hat W$ and $\hat U$.
The  $\hat W$ and $\hat U$ rotations
influence, however, the  $L$-violating vertices. At tree level
they  become
\be
\lambda_{233}''
[S^c (\hat W D')_3 (\hat U \nu)_3
-S^c (\hat W U)_3 (\hat U E)_3
- B^c (\hat W D')_3 (\hat U \nu)_2
+ B^c (\hat W U)_3 (\hat U E)_2].
\l{mix-superpot}
\ee
As we discussed before,
in the case of $(B+L)$ conserving modes
the squark $\tilde{s}^c$ emitted from the $B$-violating
vertex mixes with
$\tilde{s},$
and the latter is absorbed in the $L$-violating vertex
(similarly, for
$\tilde{b}^c$).
According to  \refe{mix-superpot}
the amplitudes of the absorption of $\tilde s$ and
$\tilde b$
are proportional to $\langle s| (\hat W d')_3 \rangle$ and
$\langle b| (\hat W d')_3 \rangle$
respectively. Thus choosing $\hat W$ in such a way that
$(\hat W d')_3 = d$ and suggesting that there is no
flavor squark mixing ({\it e.g.}\ $\tilde{s}^c$ and $\tilde{d}$)
one can suppress all $(B+L)$
decay modes in one-loop. Similar consideration holds for box diagrams.

Let us note that in the case of strong mixing effect or 
permutation, $\hat W$, 
family structure itself and family hierarchy of $R$-parity violating  
couplings has no sense. There is no  
at least strong correlation of the couplings with fermion masses. 
\vskip0.5truecm

The propagation  of $\tilde{s}^c$ as well as $\tilde{b}^c$
between the vertices \refe{B-vertex} and \refe{mix-superpot}
results in the following $(B-L)$-conserving operators:
\be
u d\  \left[
(\hat W d')_3 (\hat U \nu)_3 -
(\hat W u)_3 (\hat U e)_3], \ \ \
u d \ [-(\hat W d')_3 (\hat U \nu)_2 +
(\hat W u)_3 (\hat U e)_2 \right].
\l{mix-operators}
\ee
Since the neutrinos are massless (or very light) the only possibility to
suppress the neutrino modes is to take
$(\hat W d')_3 \equiv b$.
Evidently, in this case the $(B+L)$-conserving modes
are unsuppressed.  Moreover, the equality  $(\hat W d')_3 \equiv b$
means that
$\langle (\hat W u)_3| u \rangle \equiv V_{ub}$ and the latter is
non-zero.
Consequently,  second and fourth terms in \refe{mix-operators}
are not removed.
Either $(\hat U e)_2$ or $(\hat U e)_3$ have an  admixture of
$e$ or $\mu,$ and from \refe{mix-operators}
one gets, for instance, the operator
$u d u \mu$ which leads to the proton decay.

Thus  the additional rotations $\hat W$ and $\hat U$ 
do not allow to remove
$(B-L)$ modes completely, but they change  branching ratios,
suppressing, {\it e.g.}, \ the neutrino modes.
Eliminating the leading $(B+L)$ modes the $\hat W$ and $\hat U$
rotations
relax the bound on $\Lambda_{233}$ by factor
$V_{ts}^{-1/2} \sim 5$.

Since CKM-mixing breaks any family
symmetry, it is impossible to suppress the proton decay completely in
the high orders of the perturbation theory. No
horizontal symmetry can be introduced to forbid the operators
of the type
$\bar u^c \bar d^c d \nu .$
$B$- and $L$-violation at least in some sector of the model will
be propagated due to CKM-mixing to operators with light fermions which
induce proton decay.\vskip0.5truecm

There are two evident possibilities to suppress proton decay:\\
1.  suppress the mixing between matter generations;\\
2.  modify the relation \refe{lambda-unified2}
between $B$- and $L$-violating couplings of
usual matter fields in such a way that either $B$- or $L$- violating
couplings are strongly suppressed ($B$-, $L$- violation asymmetry).

In the first case (since the mixing of  known fermions is determined)
one should introduce fourth fermion family,
$\bar{5}_4$, $10_4$ that
has very small mixing with other families.
For instance, the  $R$-parity violating coupling
$\Lambda_{234} \bar{5}_2 \bar{5}_3 10_4$
generates the neutrino mass in the cosmologically interesting 
region ($\sim$ 10 eV) but does not result in fast proton decay, 
if the mixing with other generations is smaller than 
$V_{Td}, V_{Ts} < 10^{-8}$. Indeed, performing 
analysis  similar to that of the sect.\ 4 one will get similar 
suppression factors with substitution: 
$t \rightarrow T$: $V_{td} \rightarrow V_{Td}$,  {\em etc.}. 
The $B$- and $L$-violating operators which are
generated by $\Lambda_{234}$ at tree level are 
$b^c T^c \nu B'$. In the mass diagonal basis 
the reduction of generation index can be done by interaction 
of the charged Higgs and gauge bosons. Therefore one inevitably 
gets suppression factors proportional to $V_{Td} V_{Ts}.$ 
Note that $B$-violating coupling from the above term
involves  quark of the fourth generation: $b^c s^c T^c$.
To get the $B$-violating coupling $b^c s^c t^c$ without 
proton decay one should
permute the fermions in multiplets in such a way that in $10_4$ the
upper quark $T^c$ is substituted by  $t^c$ and
in  $\bar{5}_3$ the lepton doublet  $l_3$ is substituted by $l_4$.

Let us note that in the case of four generations 
$V_{td}$ can be zero and according to \refe{cons-bound}
proton decay at  one-loop level disappears.
However, since $V_{ts}$ is non zero the one-loop diagrams 
will generate operators $u s \bar{s^c} \bar{\nu}$ 
or $ u s s \nu $ which can be converted into 
operators inducing proton decay by additional $W$ exchange. 
Thus proton decay appears in two-loops. Additional 
suppression factor is $g^2/16 \pi^2\ V_{us} \approx 10^{-3}$. 
This in turn relax the bound on $\Lambda$ by 1.5 order 
of magnitude.\vskip0.5truecm

Concerning the second possibility, let us note that
in Grand Unified
theory with quark and lepton unification
it is nontrivial to get the $B$- and $L$-violation
asymmetry.
As we will see in the next section the asymmetry could 
be related to the doublet-triplet splitting.

\section{$R$-parity violation and doublet-triplet splitting.}

There are two possible ways to relate the
asymmetry between the $L$- and $B$-violating couplings
of usual matter fields in GU theories
with doublet-triplet mass splitting.

1. Due to mixing of the matter and Higgs 5-plets
(second term in \refe{R-viol-su5})
the doublet-triplet splitting of the
Higgs multiplet can lead to doublet-triplet asymmetry of
matter field multiplet.
This in turn breaks symmetry between quarks and
leptons, and eventually,  between the $L$- and $B$-violating couplings.
Such a situation is realized in the model by Hall and Suzuki
\cite{Hall-Suzuki}.

Let us consider an example of model, where the
matter-Higgs mixing is  the only source of
$R$-parity violation.  Suggesting as before that the
third generation coupling dominates,  we can
write the appropriate terms of
the superpotential in the following way
\be
\bar 5_3 \hat m H +
\bar H \hat M H + y_{i}\ \bar 5_i 10_i \bar H ,
\l{sup}
\ee
where $\bar 5_i$ and $10_i$ are defined in the diagonal basis for down quark
Yukawa couplings $y_i$, $i=d,s,b$
so that $d^c_i$ and $d_i$ coincide,
up to corrections $M_W/M_\GU,$
with mass eigenstates. 
The mass matrices  of \refe{sup}
can be written in the doublet-triplet form as:
\be
\hat m = {\rm diag}(m_\T, m_\D), \ \ \ \hat M = {\rm diag}(M_\T, M_\D),
\l{matr}
\ee
where $M_\T \sim M_\GU$ and $M_\D$, $m_\D$ and $m_\T$
are at the electroweak
scale (large value of  $m_\T$ would result in the fast proton decay).
The first term in \refe{sup} can be eliminated by
rotations of the doublet and the triplet components of the
5-plets: $\bar 5_3=(B^c,L_3)$ and $\bar H=(\bar {\cal T}, H_1)$.
For triplet components we get the mixing:
\be
\begin{array}{lcl}
\bar {\cal T}'  &=& c_\T \bar {\cal T} + s_\T B^c\\
{B^c}'  &=& c_\T B^c - s_\T \bar {\cal T},
\end{array}
\l{trip}
\ee
where ${B^c}'$ and $\bar {\cal T}'$ are the mass states,
$c_\T \equiv \cos \theta_\T$,
$s_\T \equiv \sin \theta_\T,$  and
\be
\frac{s_\T}{c_\T} = \frac {m_\T}{M_\T}.
\l{trip-mix}
\ee
For doublet components:
\be
\begin{array}{lcl}
H_1' &=& c_\D H_1 + s_\D L_3\\
L_3' &=& c_\D L_3 - s_\D H_1,
\end{array}
\l{doub}
\ee
and
\be
\frac{s_\D}{c_\D} = \frac {m_\D}{M_\D}.
\l{doub-mix}
\ee
Since $m_\D, m_\T, M_\D\sim M_W$ one gets
from \refe{doub-mix}
and \refe{trip-mix} that
$s_\T$
is strongly suppressed, $s_\T \sim M_W/ M_\GU < 10^{-14}$, 
whereas $s_\D$ can be of the order 1.

Substituting the expressions \refe{trip} and \refe{doub} into 
\refe{sup} we obtain the effective $R$-parity violating
couplings \refe{R-parity-violating}.
In particular the third generation Yukawa coupling gives
\be
\leff L_3 B^c Q_3',
\l{lambda333}
\ee
where
\be
\leff = s_\D\cdot y_b,
\ee
and $Q_3' \equiv V_{ib}^* Q_i$. Baryon violating interactions as well
as pure leptonic terms are absent due to the antisymmetry.
The Yukawa coupling of the  second generation leads to
\be
y_s\ 
[ s_\T  B^c S^c U^c_i +
  s_\D  L_3 S^c Q_i   +
  s_\D  L_2 L_3 E^c_i
]
\l{sec}
\ee
(The first generation Yukawa coupling gives similar terms with
substitution $y_s V_{is}\to y_d V_{id},$
$S \to D$, $L_2 \to L_1$).

The leading contribution to the proton decay is induced by
$L$-violating interaction
\refe{lambda333} and $B$-violating interaction \refe{sec}. The
$\tilde{b}^c$ exchange dressed by  $H^+$, $\tilde{H}^+$...
results in
the amplitude for proton decay
\be
A \propto
\leff \cdot y_s\ s_\T\cdot \xi_{B-L}
= y_s y_b\ s_\D s_\T\ \xi_{B-L}.
\ee
Substituting values of parameters,  we find that even for large $\tan
\beta$
($y_b \sim 1$) this amplitude is small enough to allow for
$s_\D$, and consequently,  $\leff$
to be of the order 1. All other
diagrams give
smaller contributions. (Note that in the considered example all
the $B$-violating interactions contain $b^c$ quark, so that
even lowest family couplings need a loop ``dressing").


There is another consequence of the 
matter-Higgs mixing \cite{banks,nir}:
Explicit $R$-parity violating terms in
\refe{sup} induces in general VEV 
of sneutrino\footnote{We are grateful 
to referee who pointed on this possibility.}.
Indeed, the relevant terms in the potential at the electroweak
scale are:
\be
\ba{rl}
V \ni & (m_{L_3}^2+\delta m^2)\ |H_1|^2 + 
m_{L_3}^2\ |L_3|^2\ - \\[1ex] \nonumber
&[B\cdot M_\D\ H_1 H_2 + (B+\delta B)\cdot m_\D\ 
\tilde L_3 H_2 + {\rm h.c.}].
\l{scalar-potential}
\ea
\ee
We suggest that soft breaking terms are universal
at a certain scale $M_X$, say the one suggested 
by gauge coupling unification
or the Planck scale.
Then the parameters $\delta m^2$ and $\delta B$ 
\refe{scalar-potential}
describe the renormalization effect due to the bottom
Yukawa coupling from $M_X$ to the electroweak scale.
The corresponding renormalization group equations are:
\be
\ba{cl}\displaystyle
\frac{d}{dt} \delta B =&  
3\ y_b^2\ A_b, 
\\[1.5ex] \nonumber \displaystyle
\frac{d}{dt} \delta m^2 =& 
3\ y_b^2\ (m_{Q_3}^2+m_{D^c_3}^2+m_{H_1}^2+A_b^2), 
\l{ren-group-for-breakers}
\ea
\ee 
where $t=1/(4\pi)^2 \times \log(M_\GU^2/Q^2).$
The rotation \refe{doub} 
which eliminates matter-Higgs mixing 
term in the superpotential generates mixing terms
for sleptons:
\be
V_\L \approx  \theta_\D \times 
\left[ 
\delta m^2\ H_1^* + \delta B\cdot \mu\ H_2 
\right] \tilde L_3 + 
{\rm h.c.}
\l{lepton-violating-part}
\ee
(for small $\theta_\D$).
After electroweak symmetry breaking 
these mixing terms, together with soft 
symmetry breaking masses, 
induce a VEV of tau sneutrino of the order:
\be
\nuvev
\sim v\ \theta_\D\times  
\left(\frac{\delta m^2}{m_{L_3}^2}\ \cos\beta +
\frac{\delta B\cdot \mu}{\; m_{L_3}^2}\ \sin\beta \right).
\l{tau-sneutrino-vev}
\ee 
The factor in brackets can be estimated as
$y_b^2\ $ $(3\ \cos\beta + 0.5\ \mu/m_{L_3}\ \sin\beta),$ 
where the figures quoted 
arise from approximate integration of renormalization group equations
\refe{ren-group-for-breakers}. Consequently
the tau sneutrino  
VEV is\footnote{ 
Technically it is possible to implement
a cancellation between the two terms in \refe{tau-sneutrino-vev} 
(see \cite{lee} for a phenomenological
study of such a possibility). However 
we see no natural reason for this to 
happen in the supergravity context.} 
$\nuvev\sim v\ \theta_\D\ y_b^2$. 
Due to this VEV the tau neutrino mixes with the zino,
and consequently the mass of tau neutrino is generated 
via the see-saw mechanism:
\be
\frac{g_1^2+ g_2^2}{2}\ \frac{\nuvev^2}{M_{\tilde Z}}
\l{neutrino-mass}
\ee
(see \cite{Hall-Suzuki,bhh}).
In the model under consideration
this contribution to tau neutrino mass is
typically larger 
than the one produced by the loop-diagram stipulated by
interaction \refe{lambda333}. 

We can derive from \refe{neutrino-mass} the  bound on 
$R$-parity violating couplings. Taking into account that  
$\leff\sim \theta_\D\ y_b,$ and  
$\nuvev \sim v\ \theta_\D\ y_b^2$ we get the relation between 
$\leff$ and neutrino mass 
\be
\leff\sim 0.06\times 
\left[\frac{\theta_\D}{0.1\ {\rm rad.}} \right]^{1/2}
\!\!\!\times \left[\frac{m_{\nu_\tau}}{10\ {\rm MeV}} \right]^{1/4}
\!\!\!\times \left[\frac{M_{\tilde Z}}{1\ {\rm TeV}} \right]^{1/4}.
\ee
Therefore it is possible to 
obtain large $R$-parity violating couplings
with tau neutrino masses close to the 
present experimental limit.


2. Another possibility to get the asymmetry of the $B$- and $L$-violation
is to introduce the
explicit doublet-triplet splitting in the matter multiplets.
For this one should assume the existence of new superheavy matter fields.

Suppose that each generation of matter field contains
an additional pair of 5-plets: $5'$ and $\bar 5'$ with doublet-triplet
splitting. For the third generation we introduce:
\be
\bar 5_3=\left(
         \begin{array}{c}
         B^c\\
         L_{3 G}
         \end{array}
         \right)\ \ \ \
\bar 5_3'=\left(
         \begin{array}{c}
         B^c_G\\
         L_3
         \end{array}
         \right)\ \ \ \
     5_3'=\left(
         \begin{array}{c}
         B_G \\
         L_{3 G}^c
         \end{array}
         \right) ,
\l{22}
\l{univ-d-t}
\ee
where $B^c_G, B_G, L_{3 G}^c,$ and $L_{3 G}$
are new superheavy
fields with mass $\sim M_\GU.$

Note that
by \refe{univ-d-t} we {\em generalize the doublet-triplet
splitting}  which is present now
not only in the Higgs multiplets but also in the matter multiplets
\footnote{
We will not discuss here the origin or the naturalness of the
doublet-triplet splitting.
Formally,
the permutation of the light and the heavy matter
fields can be achieved {\it e.g.}\ due to the interaction:
$$
\bar 5_3 (M+h \Phi) 5_3'+
\bar 5_3' (M'+h' \Phi) 5_3' ,
\l{permutation}
$$
where
the parameters $M, M', h,$ and $h'$ are adjusted in such
a way that $B^c$ and $L_3$ are massless at the GU scale,
whereas $B^c_G,B_G, L_{3 G}^c,$ and $L_{3 G}$
acquire masses $O(M_\GU).$}. This ``universal"
doublet-triplet splitting could have an unique origin.

The electroweak symmetry breaking via  the interaction
$\bar 5'_3 10_3 H$,
results in mixing of the heavy and the light component with typical
mixing angles:
\be
\tan \alpha \sim \frac{M_W}{M_\GU} \sim 10^{-14}\ .
\ee
Using the  multiplets  \refe{univ-d-t}
one can introduce $R$-parity violating
interactions even within one generation:
\be
\Lambda_{333} \bar 5_3 \bar 5_3' 10_3.
\ee
This  gives the terms:
\be
B^c B^c_G T^c+ B^c Q_3 L_3 - L_{3 G} Q_3 B^c_G + L_{3 G} L_3 \tau^c.
\l{26}
\ee
Note that there is no $B$-violating terms with only
light matter fields.
Mixing between $B^c$ and $B^c_G$ does not lead to
such a term due to the antisymmetry of interaction.
Proton decay is generated by  one-loop diagram
of the type shown in Fig. 1a with $B^c$ being substituted by $B^c_G$. The
corresponding suppression factor
\be
\xi \frac{m_{\tilde{b}}^2}{M_\GU^2} \ln \frac{m_H}{M_\GU}
\ee
is strong enough to remove the bound on $\leff$.
As in the previous case \refe{lambda333} the
only $R$-parity violating coupling of light
fields is the one with $L$-violation.
It generates the neutrino mass at one-loop 
via bottom-sbottom exchange.

Let us consider the possibility to get the  $B$-violating
coupling $s^c b^c t^c$.
For this we introduce the additional
5-plets $\bar 5'_2$ and $5'_2$ of second generation
with new superheavy
fermions $S^c_G, S_G, L_{2 G}^c,$
and $L_{2 G}$ and with permutation of light and
heavy fermions, similar to that in \refe{22}.
Now apart from the desired term
$\bar{5}_2 \bar{5}_3 10_3$  one should admit also  all other
interactions which can be obtained from this
by substitution
$\bar 5_2
\leftrightarrow
\bar 5_2'$ and
$\bar 5_3
\leftrightarrow
\bar 5_3'$:
\be
(f_{333}      \bar 5_3  \bar 5_3' +
\lambda_{233} \bar 5_2  \bar 5_3 +
f_{233}       \bar 5_2' \bar 5_3 +
f_{323}       \bar 5_2  \bar 5_3' +
g_{233}       \bar 5_2' \bar 5_3' +
f_{223}       \bar 5_2 \bar 5_2' ) 10_3.
\l{other-couplings}
\ee
(In fact,  the permutation implies that the multiplets
with permuted components have the same quantum numbers).
However, if all these terms are present at once,   
they reproduce
all the $R$-parity violating interactions
\refe{R-parity-violating} with light
matter fields,
and thus lead to the situation discussed in sect.\ 5.
One possibility to solve the problem is to suggest strong hierarchy of
couplings in \refe{other-couplings}.
Also  family symmetry can be introduced which
forbids all the terms in \refe{other-couplings}
but  the desired one.
For instance, $U(1)$ symmetry with zero charge for
$\bar{5}_2$, $\bar{5}_3$  $10_3$ and charge 1 for all the rest
multiplets makes the desired selection. However, such a
symmetry will be broken by mass terms,
although this violation does not destroy  the suppression of
proton decay.

Let us finally remark that the doublet-triplet splitting
breaks the $SU(4)$ symmetry responsible for $b-\tau$
unification at the GU scale.
For instance in the model \refe{sup} the
mass terms for bottom quark and tau lepton
appear with the same couplings:
\begin{equation}
B^c Q_3 H_1 + \tau^c L_3 H_1,
\end{equation}
but after the rotation \refe{doub} we get:
\begin{equation}
B^c Q_3 (c_\D H_1'-s_\D L_3') + \tau^c L_3' H_1'.
\end{equation}
Consequently the generation of the $R$-parity violating coupling
$L_3 B^c Q_3 $ with constant proportional to $s_\D$
turns out to be connected to  the reduction of the $b-\tau$
mass ratio by the factor $c_\D$\footnote{Another contribution
to the bottom quark mass may come from the VEV of the tau sneutrino.
We consistently neglect both effects when studying the 
tau sneutrino VEV, however they have to be considered {\em e.g.}\ 
in the study of third family Yukawa coupling unification.}.

\section{Discussion and Conclusions.}

1. The $R$-parity violating couplings may have strong
flavor hierarchy, so that the coupling constants for
the fields from the third generation could be of the order 1.
These couplings may have a number
of phenomenological consequences: generation of the MeV mass
of $\nu_{\tau}$, change of the infrared fixed point of the top quark,
renormalization of the mass ratio $m_b/m_{\tau}$, {\em etc.}.\vskip0.5truecm

2. Motivated by the success of the
supersymmetric Grand Unification, we have
considered the possibility
of existence of  such large couplings
in the Grand Unified theories.
In the lowest order
of perturbation theory the bound from the proton decay
can be satisfied by smallness or absence of
couplings for low generations.
However, being suppressed
in the lowest order
the proton decay appears
inevitably  at one-loop as the consequence of
the CKM-mixing. In  the safest case with only one nonzero coupling
$\Lambda_{233}$
the bound \refe{cons-bound}
$\Lambda_{233} \ltap 3\cdot 10^{-9}$ has been obtained,
which can be considered as the conservative bound on all $R$-parity
violating couplings in $SU(5)$ models.\vskip0.5truecm

3. The analysis and the bounds obtained here
are valid in a more general context. They correspond to
the bounds on  products of certain (see sect. 4) 
$B$- and $L$-violating couplings
$\lambda' \lambda'' \ltap 5\cdot 10^{-17}$.  
\vskip0.5truecm

4. In models with $R$-parity violation, especially in the case
of one-loop induced decay, the proton decay modes
may differ  from those in the usual supersymmetric model.
In particular, the modes with $(B+L)$-conservation,
like $p\to K^+ \nu_\mu$ and $n\to K^+ \mu^-$,
can  dominate over the $(B-L)$-conserving modes,
like $p\to K^+ \bar \nu_\tau$ and $p\to K^0 \mu^+.$\vskip0.5truecm

5. The bound \refe{cons-bound}
can be avoided if new fermions (new matter fields)
exist which mix very weakly  with known fermions. These could be
the fermions  from the fourth generation.

The bounds can also be avoided if there is an asymmetry of
$B$- and $L$-violating interactions, namely if
either $L$- or $B$-violating interactions are
strongly suppressed.
This asymmetry can be related to the doublet-triplet
splitting. In the simple examples the largest $R$-parity
violating coupling is the one with $L$-violation.
\vskip0.5truecm

6. For  coupling constants $\Lambda$ satisfying the bound
\refe{cons-bound}, no
appreciable effects of $R$-parity violation in
accelerator experiments are expected.
Also, the generated neutrino masses are very small.
Inversely,  the observation of $R$-parity violating effects at
accelerators will have strong impact on the Grand Unification:
this can imply Higgs-matter
mixing or doublet-triplet splitting in matter
supermultiplets.\vskip0.5truecm

\vskip.5truecm
\noindent{\bf \LARGE Acknowledgements.}
\vskip0.6truecm

The authors  would like to thank  B. Brahmachari,
S. Bertolini,
H. Dreiner,  G. Dvali,
E. Roulet, and J.W.F. Valle for useful discussions.
\vfill\eject
\noindent{\bf \LARGE Figure Captions}
\vskip2truecm
\noindent{\large Fig.\ 1:}
{Leading one-loop diagrams of $(B-L)$-conserving $p$-decay
in the model with $\Lambda_{233}\neq 0.$
Similar diagrams exist with $\tilde s^c$
exchange and the emission of $\bar\nu_\tau.$}
\vskip1truecm
\noindent{\large Fig.\ 2:}
{Leading one-loop diagrams of $(B+L)$-conserving
$p$-decay
in the model with $\Lambda_{233}\neq 0.$
Similar diagrams exist with substitution
$H\to W,$ $\tilde H\to \tilde W.$}
\vfill\eject


\begin{thebibliography}{99}

\bibitem{reviews}
For a general review see H.P. Nilles,
\prep{110}{84}{1};
for phenomenology see H.E. Haber and G.L. Kane,
\prep{117}{85}{75}; for searches of supersymmetric particles
ALEPH Collaboration, \prep{216}{92}{253}.


\bibitem{WSH}
$R$-parity breaking couplings have been
first considered by S. Weinberg, \prd{26}{82}{287} and
N. Sakai and T. Yanagida, \plb{197}{82}{533};
for phenomenology
see 
J. Ellis {\it et al.}, \plb{150}{85}{142};
G.G. Ross and J.W.F. Valle, \plb{151}{85}{375};
S. Dawson, \npb{261}{85}{297};
a brief review is in L.J. Hall, \mpla{7}{90}{467}.


\bibitem{Zwirner}
F. Zwirner, \plb{132}{83}{103}.

\bibitem{Barbieri-Masiero}
R. Barbieri and A. Masiero,
\npb{267}{86}{210}.

\bibitem{Goity-Sher}
$n-\bar n$-oscillations  have been recently reconsidered by
J.L. Goity and M. Sher,
\plb{346}{95}{69}.


\bibitem{HN}
I. Hinchliffe and T. Kaeding,
\prd{47}{93}{279};
Y. Nir and V. Ben-Hamo,
\plb{399}{94}{77}.

\bibitem{pd}
F. Vissani, \prd{52}{95}{4245}.


\bibitem{neutrino-masses1}
K.S. Babu and R.N. Mohapatra, \prl{64}{90}{1705};
R. Barbieri, M.M. Guzzo, A. Masiero and D. Tommasini, \plb{252}{90}{251};
E. Roulet and D. Tommasini, \plb{256}{91}{218}.

\bibitem{neutrino-masses2}
K. Enqvist, A. Masiero and A. Riotto, \npb{373}{92}{95}.

\bibitem{0nu-bb1}
R.N. Mohapatra, \prd{34}{86}{3457};
J.D. Vergados, \plb{184}{87}{55}.

\bibitem{0nu-bb2}
M. Hirsch, H.V. Klapdor-Kleingrothaus and S.G. Kovalenko,
\plb{352}{95}{1}
and preprint
{\tt hep-ph/9502385} ;
K.S. Babu and R.N. Mohapatra,
\prl{75}{95}{2276}.

\bibitem{Barger-Giudice-Han}
V. Barger, G.F. Giudice and T. Han,
\prd{40}{89}{2987}.

\bibitem{lsp-decay}
For colliders searches:
H. Dreiner and G.G. Ross, \npb{365}{91}{597};
R.M. Godbole, P. Roy and X. Tata, \npb{401}{93}{67};
L. Roszkowski, Proceedings of {\it Wailikoa 1993}, 854;
V. Barger, M.S. Berger, P. Ohmann and R.J.N. Phillips,
\prd{50}{94}{4299};
H. Baer, C. Kao and X. Tata, \prd{51}{95}{2180}.

\bibitem{CRS}
C.E. Carlson, P. Roy and M. Sher,
\plb{357}{95}{99}.

\bibitem{hierarchy}
S. Dimopoulos and L.J. Hall, \plb{207}{87}{210};
S. Dimopoulos, R. Esmailzadeh, L.J. Hall, J.-P. Merlo
and G.D. Starkman, \prd{41}{90}{2099};
H. Dreiner and G.G. Ross, \npb{365}{91}{597}.




\bibitem{Bh}
G. Bhattacharyya, J. Ellis and K. Sridhar,
\mpla{10}{95}{1583};
G. Bhattacharyya, D. Choudhury and K. Sridhar,
\plb{355}{95}{193}.


\bibitem{Biswajoy}
B. Brahmachari and P. Roy, \prd{50}{94}{39};
erratum, \prd{51}{95}{3974}.

\bibitem{prep}
A.Yu. Smirnov and F. Vissani, paper in preparation.

\bibitem{GCC1}
M.B. Einhorn and D.R.T. Jones, \npb{196}{82}{475}.

\bibitem{GCC2}
U. Amaldi, W. de Boer and H. Furstenau, \plb{260}{91}{447};
U. Amaldi \etal, \plb{281}{92}{374}.

\bibitem{GCC3}
J. Ellis, S. Kelley and D.V. Nanopoulos, \npb{373}{92}{55};
F. Anselmo, L. Cifarelli, A. Peterman and A. Zichichi,
\nc{105}{92}{1817}; \nc{105}{92}{581}; \nc{105}{92}{1201};
P. Langacker and N. Polonsky, \prd{47}{93}{4028};
M.Carena, S. Pokorski and C.E.M. Wagner, \npb{406}{93}{509};
V. Barger, M.S. Berger and P. Ohmann, \prd{47}{93}{1093}.

\bibitem{btau}
M.S. Chanowitz, J. Ellis and M.K. Gaillard, \npb{128}{77}{506};
A.J. Buras, J. Ellis, D.V. Nanopoulos and M.K. Gaillard,
\npb{135}{78}{66}.


\bibitem{btau-in-susy}
See, {\it e.g.}: H. Arason, D.J. Casta\~no, B. Kesthelyi, S.
Mikaelian, E.J. Piard, P. Ramond and
D.B. Wright, \prd{46}{92}{3945} and references
therein.

\bibitem{Ramond}
P. Ramond, Proceedings
of {\it Orbis Scientiae 1983}, 91;
M.J. Bowick, M.K. Chase and P. Ramond, \plb{128}{83}{185}.

\bibitem{Hall-Suzuki}
L.J. Hall and M. Suzuki, \npb{231}{84}{419}.

\bibitem{Brahm-Hall}
D. Brahm and L.J. Hall, \prd{40}{89}{2449}.

\bibitem{banks} 
T. Banks, Y. Grossman, E. Nardi and Y. Nir,
\prd{52}{95}{5319}.

\bibitem{nir}
N. Polonsky, talk at the
1995 Summer Institute on {\em Signals of Unified Theories,} Sept. 2-8,
Laboratori Nazionali del Gran Sasso;
R. Hempfling, preprint {\tt hep-ph/9511288.} 

\bibitem{lee} I-H. Lee, \plb{138}{84}{121}; \npb{246}{84}{120}.

\bibitem{bhh}
D.E. Brahm, L.J. Hall and S.D.H. Hsu, \prd{42}{90}{1860}.


\end{thebibliography}
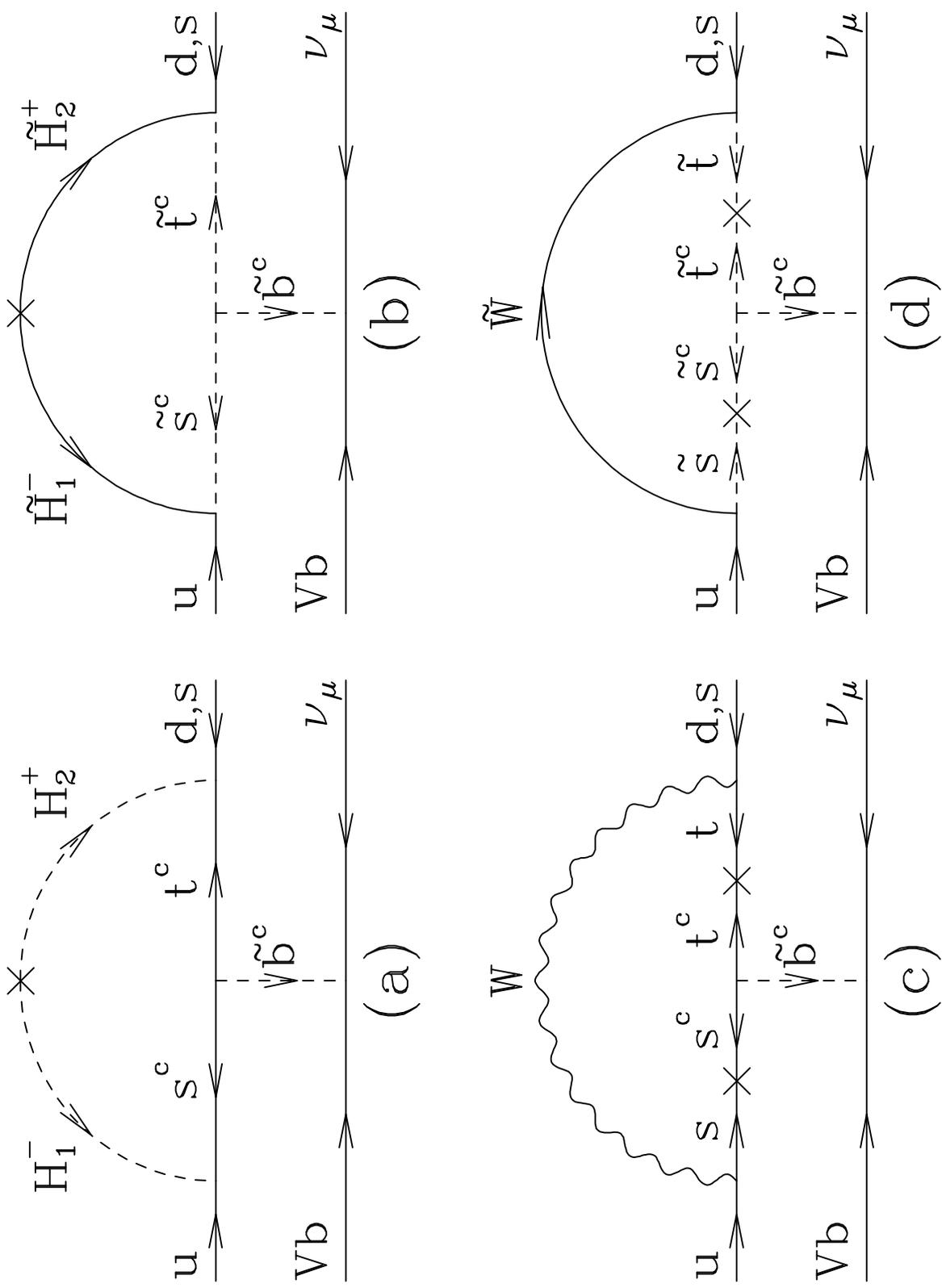
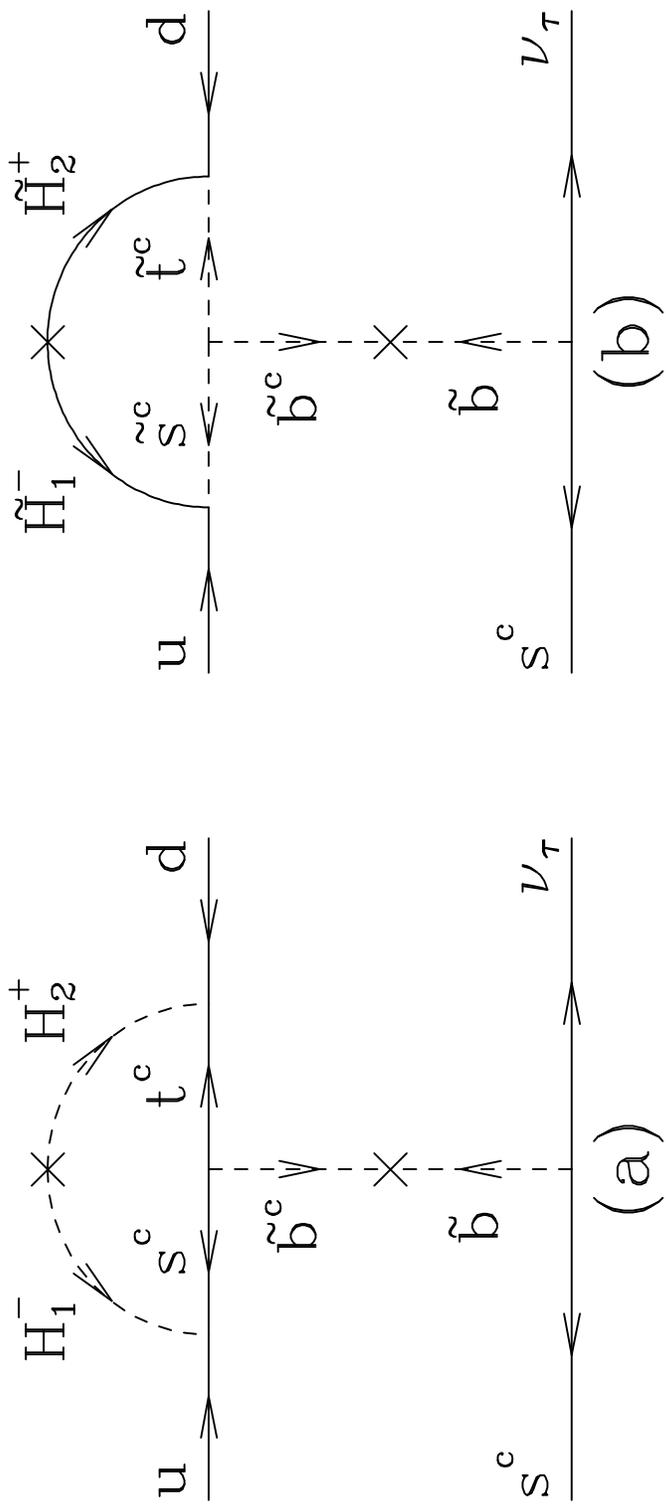
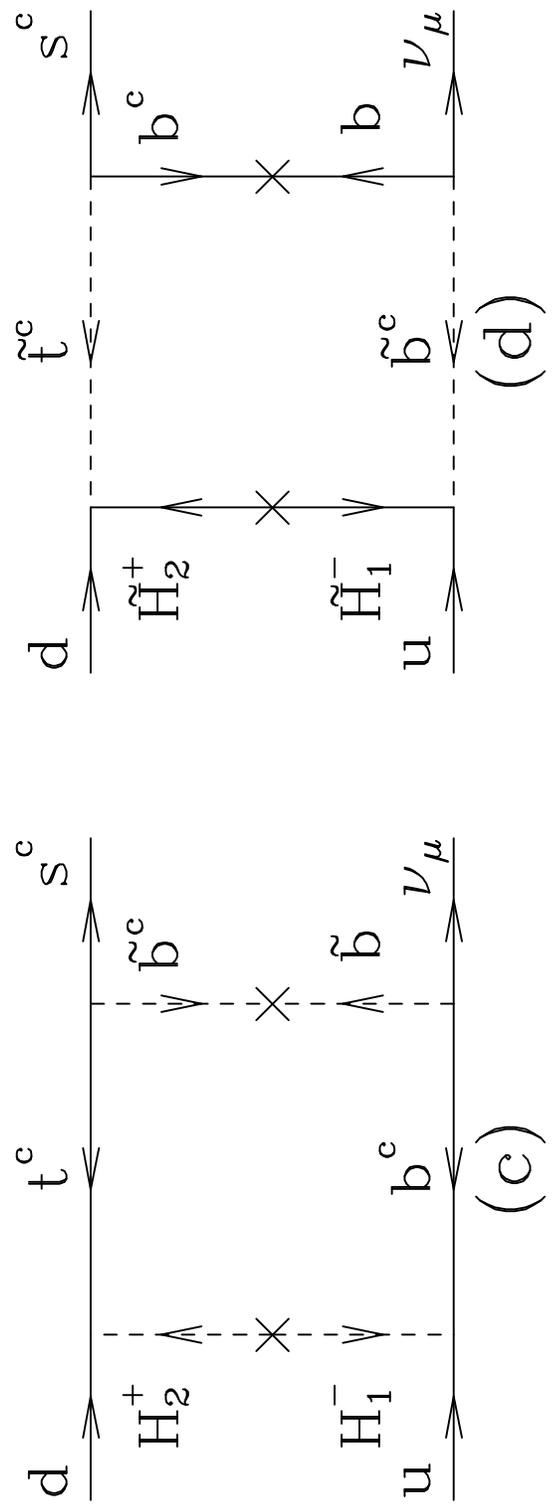
\end{document}